\documentclass[
book
]{agujournal2018}
\usepackage{apacite}
\usepackage{url} 
\usepackage{lineno}
\usepackage{soul}
\draftfalse
\usepackage{amssymb}

\newcommand{\grl}{{Geophys. Res. Lett.}}

\newcommand{\prl}   {{\it Physical Review Letters}}
\newcommand{\apj}   {{Astrophys. J.}}
\newcommand{\apjl}  {{Astrophys. J. Lett.}}
\newcommand{\mnras}   {{MNRAS}}
\newcommand{\aap}   {{Astron. Astrophys.}}

\newcommand{\ssr}{{Space Sci. Rev.}}

\journalname{Enter journal name here}

\begin{document}

%
%


\title{Tearing instability in Alfv\'en and kinetic-Alfv\'en turbulence}

%
%




\authors{Stanislav Boldyrev\affil{1,2}, Nuno F. Loureiro\affil{3}}
\affiliation{1}{Department of Physics, University of Wisconsin-Madison, Madison, WI 53706, USA}
\affiliation{2}{Space Science Institute, Boulder, Colorado 80301, USA}
\affiliation{3}{Plasma Science and Fusion Center, 
Massachusetts Institute of Technology, 
Cambridge, MA 02139, USA}





\correspondingauthor{Stanislav Boldyrev}{boldyrev@wisc.edu}





%
%


\begin{abstract}
Recently, it has been realized that magnetic plasma turbulence and magnetic field reconnection are inherently related phenomena. Turbulent fluctuations generate regions of sheared magnetic field that become unstable to the tearing instability and reconnection, thus modifying turbulence at the corresponding scales. In this contribution, we give a brief discussion of some recent results on tearing-mediated magnetic turbulence. We illustrate the main ideas of this rapidly developing field of study by concentrating on two important examples -- magnetohydrodynamic Alfv\'en turbulence and small-scale kinetic-Alfv\'en turbulence.    
Due to various potential applications of these phenomena in space physics and astrophysics,  we specifically try not to overload the text by heavy analytical derivations, but rather present a qualitative discussion accessible to a non-expert in the theories of turbulence and reconnection.   
\end{abstract}

%
%

%


%
%
%
%

\section{Introduction}
\label{introduction}
Most of our understanding of turbulence has come from everyday experience, and it is related to turbulence in fluids and gases (e.g., oceans and atmosphere). Beyond the terrestrial applications, however, one encounters turbulence that is of a qualitatively different nature. The matter is that most of the interplanetary and interstellar media are filled with plasma, an ionized gas that is a very good conductor. Plasmas can sustain electric currents that generate magnetic fields whose energy can be comparable to the kinetic energy of the plasma motion. Magnetic field and magnetic forces thus become crucial components of the plasma dynamics, which qualitatively changes the character of turbulence \cite[e.g.,][]{biskamp2003,elmegreen2004,chen2016,verscharen2019,landi2019,bruno2013,matthaeus2011,tsurutani2018}.  

Turbulent motion can contain substantial energy, which is gradually transferred from large- to small-scale fluctuations and eventually removed from turbulent motion by kinetic processes, for instance, by viscous or resistive dissipation.  The self-similar energy transfer from large to small scales, a Kolmogorov turbulent cascade, is a central concept in modern theories of turbulence \cite[e.g.,][]{frisch1995,biskamp2003}. Recently, it has been proposed that in magnetic turbulence, the energy cascade from large to small scales can be qualitatively altered {\em before} it reaches the kinetic dissipation scales \cite[][]{loureiro2017,mallet2017,boldyrev_2017}. The cascade modification occurs due to the phenomenon of magnetic reconnection, in which magnetic-field lines advected and tangled by a flow break and change their topology.  The processes of magnetic turbulence and magnetic reconnection thus turn out to be inherently related -- turbulent motions generate sheared magnetic structures where magnetic reconnection sets in, which, in turn, modifies turbulent fluctuations at the corresponding scales. Such effects are not possible in a non-conducting fluid.

Depending on the plasma parameters, one can encounter various regimes of magnetic turbulence in astrophysical systems. At scales that are much larger than the plasma microscales, such as the ion inertial length or ion gyroscale, plasma dynamics can often be described in the framework of magnetohydrodynamics (MHD).  In a collisionless plasma, on the other hand, the energy cascade can reach very small scales, smaller than the ion gyroscale or the ion inertial length, in which case more sophisticated descriptions of the plasma are needed. Recent analytical, numerical, and observational studies indicate that effects of magnetic reconnection can be present in a variety of such turbulent regimes, which makes one conjecture that reconnection is, in fact, a general phenomenon and an inherent property of magnetic turbulence \cite[][]{loureiro2017a,mallet2017a,mallet2017,loureiro2018,vech2018,comisso2018,walker2018,dong2018,boldyrev2019}. 

In the present discussion we illustrate the role of reconnection in plasma turbulence by using two important examples: resistive magnetohydrodynamics and the collisionless kinetic-Alfv\'en dynamics. In both cases, we formulate the equations describing the dynamics, discuss the corresponding phenomenological models of strong turbulence, and then demonstrate how the turbulent cascades are modified by the tearing instability.\footnote{In this work, we use the terms ``magnetic reconnection'' and ``tearing instability'' somewhat interchangeably. It is, however, important to note that the tearing instability is the initial nonlinear stage of magnetic reconnection, while the term ``reconnection'' is often used to describe a broader range of associated fully nonlinear phenomena, such as the X-point collapse, formation of plasmoid chains, etc. In a turbulent environment, an eddy susceptible to the tearing instability may or may not evolve into these nonlinear regimes. This  question was recently analyzed in more detail in \citet{loureiro2019}. The transition to these fully nonlinear regimes, therefore, is not assumed in the present work where we study the influence of the tearing instability on magnetic turbulence.} We mostly concentrate on phenomenological aspects of the theory, and we specifically emphasize the role of conservation laws in establishing the tearing-mediated regimes of turbulence. We also note that our brief discussion does not provide an overview of the (enormous) literature existing on magnetic plasma turbulence and magnetic reconnection. Our goal is rather to give a qualitative introduction to the subject, which could motivate further reading. 

We also note that in our discussion we concentrate on typical statistical properties of turbulence, the so-called mean-field theory. We do not discuss rare, intermittent events that are also present in a turbulent flow. It is well known that even steadily driven statistically homogeneous turbulence generates spatially and temporally intermittent structures that can possess relatively strong gradients of fluctuating fields. These structures can lead to localized events of energy dissipation, particle acceleration, or radiation \cite[e.g.,][]{zhdankin2012,zhdankin2017,boldyrev_yusef-zadeh2006,zheng2018,osman2011}. Moreover, in natural systems, turbulence can be driven in a non-homogeneous fashion thus creating large-scale structures not related to turbulence itself, say corresponding to boundaries between magnetic-flux tubes \cite[e.g.,][]{borovsky2008}. It is however, generally expected that intermittency effects lead to relatively small corrections to the mean properties of turbulence (energy spectrum, rate of energy dissipation, etc) \cite[e.g.,][]{frisch1995}. The discussion of intermittency effects is beyond the scope of the present work.

\section{Magnetohydrodynamic turbulence}
\label{mhd_turbulence}
As discussed above, magnetic fields play essential role in the dynamics of a conducting fluid. Magnetic fields may be produced by external sources, as it happens, for instance, in the solar corona where the magnetic loops are generated in the interior of the sun and buoy above the surface, or in laboratory devices where they are imposed by external coils. Magnetic fields can also be generated by random motion of a conducting fluid itself through the so-called magnetic dynamo action. In all cases, once the magnetic fields are generated at a certain scale, they affect the dynamics of fluctuations at all smaller scales. Indeed, while a uniform large-scale velocity field can be removed by choosing a co-moving inertial frame (a Galilean transform), the uniform magnetic field cannot. In the discussion of magnetic turbulence, it is, therefore, important to assume the presence of a uniform magnetic field. We will choose the coordinate frame in such a way that this magnetic field is in the $z$-direction, and denote it as ${\bf B}_0$. 

In our discussion of MHD turbulence we will also assume that the fluid motion is incompressible. This is a good approximation since in the presence of a uniform magnetic field, magnetohydrodynamic fluctuations are typically dominated by shear-Alfv\'en modes that are incompressible and that correspond to shuffling of magnetic field lines rather than to bending or stretching them. This is also consistent with observations; for example, plasma turbulence in the warm interstellar medium is characterized by Mach numbers on the order of, or smaller than, one, \cite[e.g.,][]{elmegreen2004} while turbulence in the solar wind leads to plasma density fluctuations that much smaller than the background density \cite[e.g.,][]{chen2016}. Therefore, we assume that the plasma density, $\rho_0$, is uniform and constant, and we will measure the magnetic field ${\bf B}$ in units of the corresponding Alfv\'en speed, $v_A=B/\sqrt{4\pi\rho_0}$. The resulting equations for the incompressible magnetic and velocity fields then read:    
 \begin{eqnarray}
&\partial_t {\bf v} + \left({\bf v} \cdot \nabla\right) {\bf v} =  - \nabla p + \left({\bf \nabla \times B}\right) \times {\bf B} + \nu \nabla^2 {\bf v}, \label{eq_mom} \\
&\partial_t {\bf B} = \nabla \times \left( {\bf v} \times {\bf B} \right) + \eta \nabla^2 {\bf B}, \label{eq_ind}
\end{eqnarray}
where $p$ is the pressure whose role is to ensure solenoidality of the velocity field, $\nu$ is the fluid viscosity and $\eta$ is the magnetic diffusivity (related to the ohmic resistivity of the fluid). Equation~(\ref{eq_mom}) is the momentum equation, while Eq.~(\ref{eq_ind}), describing the evolution of the magnetic field, is the induction equation. 

An important role in the MHD dynamics is played by the conservation laws. In the absence of external forcing and dissipation (expressed by the viscous and diffusive terms), the MHD equations~(\ref{eq_mom}, \ref{eq_ind}) conserve the following quadratic integrals: the energy 
\begin{equation}
E=\frac{1}{2}\int \left( v^2+B^2 \right)\,d^3 x,
\label{def_energy}
\end{equation}
 the cross-helicity 
\begin{equation}
H^C=\int \left({\bf v} \cdot {\bf B}\right)\,d^3 x,
\label{def_cross}
\end{equation}
 and the magnetic helicity
\begin{equation}
H^M=\int \left({\bf A} \cdot {\bf B}\right)\,d^3x,
\label{def_maghel}
\end{equation}
where ${\bf A}$ is the vector potential, ${\bf B} = \nabla \times
{\bf A}$ \cite[see, e.g.,][]{biskamp2003,tobias2013}. 

Analysis of these integrals helps to elucidate some general properties of the MHD turbulent cascade. For instance, one can argue that in decaying MHD turbulence (that is, turbulence without a driving force) the energy transfer in Fourier space tends to be directed from modes with small wavenumbers $k$ to the modes with large wavenumbers, the so-called direct cascade. The cascade of magnetic helicity, however, proceeds in the opposite direction, from large to small wavenumbers, the so-called inverse cascade. Since the dissipation terms are most efficient at large wavenumbers, the magnetic helicity is better conserved than the energy. The resulting selective decay favors formation of large-scale helical magnetic structures \cite[e.g.,][]{pouquet1976,biskamp2003}. The role of the cross-helicity is more subtle. Since it has the same dimension as the energy it also cascades toward large wavenumbers; however, in the presence of viscosity  or magnetic diffusivity,  cross-helicity does not have sign-definite dissipation. Therefore, in general, it may dissipate slower than the energy, thus leading to the creation of so-called dynamically aligned structures, where the magnetic and velocity fields tend to align or counter-align their directions \cite[e.g.,][]{biskamp2003,boldyrev_spectrum_2006}. 

In our present discussion of magnetic plasma turbulence, in both Alfv\'en and kinetic-Alfv\'en regimes, we would like to draw attention to yet another possible role of the conservation laws: their role in the formation of sheared magnetic structures -- current sheets -- in steady-state magnetic turbulence, which may lead to the onset of tearing instability. We believe that magnetic turbulence is qualitatively different from ordinary non-conducting turbulence in this respect. For instance, while ordinary turbulence described by the Navier-Stokes equation may also possess two integrals of motion that have different cascade rates (say energy and enstrophy in the 2D case \cite[e.g.,][]{biskamp2003}), the formation of sheared velocity profiles is impeded in such cases by the Kelvin-Helmholz (KH) instability. The presence of a relatively strong sheared magnetic field qualitatively changes the picture. It may stabilize such profiles to the KH mode; however, it makes them unstable to the tearing mode instead.  

In order to analyze MHD turbulence, let us separate the constant uniform magnetic field from the fluctuating magnetic field, ${\bf B}({\bf x}, t)={\bf B}_0+{\bf b}({\bf x}, t)$. The equations for magnetic and velocity fluctuations can be written in an especially useful form if one introduces the Els\"asser variables, ${\bf z}^\pm={\bf v}\pm {\bf b}$,
\begin{equation}
  \left(\frac{\partial}{\partial t} \mp {\bf v}_A\cdot\nabla\right){\bf 
  z}^\pm + \left({\bf z}^\mp\cdot\nabla\right){\bf z}^\pm = -\nabla p + \frac{1}{2}(\nu +\eta)\nabla^2{\bf z}^{\pm}+\frac{1}{2}(\nu -\eta)\nabla^2{\bf z}^{\mp}, 
  \label{mhd-elsasser}
\end{equation}
where ${\bf v}_A$ is calculated using the background magnetic field ${\bf B}_0$. In many practical systems, turbulent magnetic fluctuations cover a broad range of scales spanning at least several orders of magnitude. In this situation, strong magnetic fluctuations at the largest scales play the role of an almost uniform guide field for the weaker small-scale fluctuations. 

At small scales, MHD turbulence becomes strongly anisotropic. Turbulent eddies are stretched along the background magnetic field so that their field-perpendicular scales are much smaller than their field-parallel scales $k_\|\ll k_\perp$. As was pointed out in \cite[e.g.,][]{goldreich_toward_1995}, this allows one to simplify the MHD equations describing small-scale fluctuation, keeping only the leading terms in the small parameter $k_\|/k_\perp$. The resulting system coincides with the so-called reduced MHD systems, also considered in \citet{kadomtsev_p74,strauss_nonlinear_1976}, and applied in many analytical and numerical studies of field-guided MHD turbulence \cite[e.g.,][]{dmitruk2003,gomez2005,rappazzo2007,perez_b2008}.

The assumptions used to derive the reduced MHD system are the following. First, one notices that for strongly anisotropic fluctuations, the principal contribution to the dynamics is given by the shear-Alfv\'en modes where the polarizations of ${\bf v}$ and ${\bf b}$ fields (or, equivalently, of ${\bf z}^+$ and ${\bf z}^-$) are normal to the background field ${\bf B}_0$. Second, one assumes that the resulting turbulence reaches the state of the so-called critical balance \cite[e.g.,][]{goldreich_toward_1995}, where the linear Alfv\'enic advection terms in Eq.~(\ref{mhd-elsasser}) become comparable to the terms describing nonlinear interaction,  $\left({\bf v}_A\cdot\nabla\right){\bf z}^\pm \sim \left({\bf z}^\mp\cdot\nabla\right){\bf z}^\pm $. We can, therefore, retain only two components of the fluctuating fields, ${\tilde {\bf z}}^{\pm}=\left({\tilde z}^{\pm}_1, {\tilde z}^{\pm}_2, 0 \right)$, however, we assume that each of those fields is three dimensional, that is, depends on three spatial coordinates. The resulting reduced MHD system then takes the form \cite[e.g.,][]{galtier2006,perez_b2008}:
\begin{eqnarray}   
\left( \frac{\partial}{\partial t} \mp {\bf v}_A\cdot \nabla \right) {\tilde
    {\bf z}}^{\pm}+({\tilde {\bf z}}^{\mp}\cdot \nabla){\tilde {\bf
      z}}^{\pm}=-\nabla_{\perp} p + \frac{1}{2}(\nu +\eta)\nabla_\perp^2{\tilde {\bf z}}^{\pm}+\frac{1}{2}(\nu -\eta)\nabla_\perp^2{\tilde{\bf z}}^{\mp} ,
\label{rmhd} 
\end{eqnarray}
Yet another useful (and, possibly, more familiar) form of the reduced MHD equations can be obtained if we represent the incompressible velocity and magnetic fields through their scalar potentials (or stream functions)  as follows, ${\bf v}={\hat{z}}\times \nabla \phi$, ${\bf b}={\hat {z}}\times \nabla \psi$. By taking curl of the momentum equation and ``uncurling'' the induction equation in the MHD system~(\ref{eq_mom}, \ref{eq_ind}),  we obtain 
\begin{eqnarray}
&\partial_t \nabla_\perp^2\phi + \left[\left({\hat{z}}\times \nabla \phi\right) \cdot \nabla\right] \nabla_\perp^2\phi -\left[\left({\hat {z}}\times \nabla \psi\right)\cdot \nabla\right]\nabla_\perp^2\psi =v_A\partial_z \nabla_\perp^2\psi +\nu \nabla_\perp^4\phi, \nonumber \\
&\partial_t \psi +\left[\left({\hat{z}}\times \nabla \phi\right)\cdot \nabla\right] \psi=v_A \partial_z \phi +\eta \nabla_\perp^2 \psi.
\label{rmhd2}
\end{eqnarray}
Based on these equations, we can now discuss the properties of magnetohydrodynamic turbulence. As we mentioned above, in a steady state, there is a flux of energy from the scales where the turbulence is excited (say, by some instability) to progressively smaller scales until small enough scales are reached where the energy is converted into heat (the viscous or resistive scales). In the interval of scales between the forcing and the dissipation, the so-called inertial interval, the rate of energy cascade over scales is constant. This rate can be estimated from Eq.~(\ref{rmhd}) if we notice that the Fourier energy density corresponding to the integral (\ref{def_energy}) is $E_k\propto |{\tilde z}^+_k|^2+|{\tilde z}^-_k|^2$. 

In our qualitative discussion, we will consider the situation when the values of the Elsasser fields $z^+$ and $z^-$ are on the same order, which is also called the case of ``balanced'' MHD  turbulence. The name comes from the fact that the Elsasser fields are closely related to the Alfv\'en waves -- the exact nonlinear solutions of the incompressible MHD equations.  Indeed, the solution of Eq.~(\ref{mhd-elsasser}) corresponding to $z^+\equiv 0$ and $z^-\neq 0$ is an exact solution of the MHD equations, the nonlinear Alfv\'en mode that propagates with the Alfv\'en velocity without distortion along the background magnetic field~${\bf B}_0$. The solution corresponding to $z^-\equiv 0$ and $z^+\neq 0$ is also exact but such a mode propagates in the opposite direction. One may expect that balanced ($z^+\approx z^-$) and imbalanced ($z^\pm > z^\mp$) dynamics are qualitatively similar \cite[see, e.g.,][]{perez_b2010,tobias2013}, so we will not dwell upon their differences here. In our dimensional analysis we will, therefore, not distinguish between ${\tilde z}^+$ and ${\tilde z}^-$ and denote them simply as~$z$. 

Dimensionally, the Elsasser and physical fluctuations are similar: $z_k\sim b_k\sim v_k$. Instead of their Fourier components, is it also useful to consider the typical fluctuations of those fields at a scale $a\sim 1/k$, which are defined as the rms values of their fluctuations over a scale $a$; for instance, $z_a \equiv \left[{\bf z}({\bf x}+{\bf a})-{\bf z}({\bf x}) \right]_{rms}$. The fluctuation energy corresponding to the scale $a$ is $E_a \propto z_a^2$. The rate of the energy cascade due to the nonlinear interaction can be estimated from Eq.~(\ref{rmhd}). As was proposed in \cite[][]{goldreich_toward_1995}, this rate may be estimated as $\gamma_a \propto z_a/a$, where $a$ is the scale of the fluctuations in the field-perpendicular direction. From the requirement that the energy flux $\varepsilon \sim E_a\gamma_a $ is constant, one derives $z_a \propto a^{1/3}$, which in the Fourier space gives 
\begin{eqnarray}
E(k_\perp)2\pi k_\perp d k_\perp \propto k_{\perp}^{-5/3} d k_\perp.\label{gs_spectrum}
\end{eqnarray} 
This is the Goldreich-Sridhar spectrum of MHD turbulence. As we mentioned above, in MHD turbulence the fluctuations are anisotropic with respect to the local guide field, meaning that $k_\perp \gg k_\|$. The energy spectrum is, therefore, the spectrum with respect to $k_\perp$, and it corresponds to the turbulent fluctuations that are Fourier-transformed only in the plane perpendicular to the guide field ${\bf B}_0$. The three-dimensional structure of the fluctuations can be described by the correlation length of the fluctuations in the field-parallel direction. Namely, fluctuations separated by a distance $a$ in the field-perpendicular direction, are correlated at a different distance $l$ along the local total magnetic field. In the Goldreich-Sridhar picture, these scales are related as $l\propto a^{2/3}$. 

The Goldreich-Sridhar model, however, does not seem to predict the spectrum of MHD turbulence correctly. Numerical simulations of incompressible turbulence with a strong uniform background magnetic field consistently reproduce the spectral exponent closer to $-3/2$ \cite[][]{muller_g05,maron_g01,haugen_04,mininni_p2007,chen10,mason_cb08,perez_etal2012,chandran_14,perez_etal2014}. This exponent may be accounted for by the phenomenon of dynamic alignment, which is related to the fact that the cross-helicity is conserved better than the energy \cite[e.g.,][]{biskamp2003,boldyrev_spectrum_2006}.
Let us discuss this in more detail, as this phenomenon will help us to understand why magnetic tearing effects may become important in MHD turbulence. The ideal reduced MHD system~(\ref{rmhd2}) has only two conserved quadratic quantities, the energy and the cross-helicity,
\begin{eqnarray}
E=\frac{1}{2}\int\left[\left(\nabla_\perp\phi\right)^2 +\left(\nabla_\perp\psi\right)^2\right]d^3 x,\\
H^C=\int\left(\nabla_\perp\phi\cdot\nabla_\perp\psi\right)d^3x.
\end{eqnarray}
Assume now that we consider a decaying (unforced) turbulent state where turbulent  fluctuations are dissipated by small viscosity and resistivity. As we mentioned above, the energy has a tendency to get dissipated faster than the cross helicity. We can therefore ask what field configurations would minimize the energy at given cross-helicity. For that we need to minimize the functional $W=E-\mu H^C$, where $\mu$ is a Lagrange multiplier. The minimization gives $\phi=\psi$ or $\phi=-\psi$ (and $\mu^2=1$), which means that the magnetic and velocity fluctuations tend to either align or counteralign their directions, leading to the so-called  Alfv\'enic states characterized by~${\bf v}({\bf x})=\pm {\bf b}({\bf x})$. This phenomenon is called the dynamic alignment. 

It is important to note that such Alfv\'enic states correspond to vanishing nonlinear terms in the MHD equations~(\ref{rmhd2}). However, in a steady turbulent state, the nonlinearities cannot vanish as they are responsible for the constant energy flux over scales. Therefore, the Alfv\'enic states cannot be achieved in steady-state turbulence. We may however assume that while the nonlinear interaction cannot vanish, the presence of the two conserved integrals may still lead to an approximate alignment of the magnetic and velocity fields, and to a depletion of nonlinear interaction. One way of achieving this is to assume that a turbulent flow creates sheared magnetic and velocity structures in the guide-field-perpendicular direction, which contain most of the energy. The large-scale magnetic field is aligned with the $z$ direction, so the structures are created in the $x-y$ plane. Assume that the magnetic and velocity fluctuations in such structures are both aligned around, say, the $y$ direction (and, therefore, their directions are approximately aligned with each other), while their magnitudes vary significantly in the $x$ direction. Such structures look like three-dimensionally anisotropic ribbons stretched along the large-scale field. Their thickness in the $x$ direction is $a$, their width in the $y$ direction is $\xi$, and their length in the $z$ direction is~$l$.  

The nonlinear interaction terms will be reduced in such sheared structures compared to the naive dimensional estimates.  For example, in Eq.~(\ref{rmhd}) one estimates for the nonlinear term at the scale $a$: $({\tilde {\bf z}}^{\mp}\cdot \nabla){\tilde {\bf z}}^{\pm} \sim (a/\xi) z_a^2 /a$. Indeed, the fields are aligned along the $y$ direction while their gradients are directed in the $x$ direction, so their dot product is reduced by the anisotropy factor~$a/\xi$. In order to reproduce the energy spectrum $-3/2$, the eddies should have the scale-dependent anisotropy  satisfying $a/\xi\sim (a/L)^{1/4}$, where $L$ is the outer scale of turbulence, that is, the scale where the energy is supplied to the fluctuations. As argued in \cite{boldyrev_spectrum_2006}, such anisotropy should indeed be expected in MHD turbulence. It is easy to demonstrate that the scale-dependent anisotropy of this type leads to the required spectrum of turbulence. Indeed, from the requirement of constant energy flux, $(a/L)^{1/4}z_a^3/a=\varepsilon$, we derive that the turbulent fluctuations scale as 
\begin{eqnarray}
b_a\sim v_a\sim \varepsilon^{1/3}a^{1/3}\left(L/a \right)^{1/12} \propto a^{1/4},
\end{eqnarray} 
and, therefore, the Fourier energy spectrum scales as $k_\perp^{-3/2}$.  
The anisotropy of turbulent eddies in the guide-field perpendicular direction consistent with $a/\xi\sim (a/L)^{1/4}$ has been detected in numerical simulations and observations \cite[e.g.,][]{mason_cb08,chen10}. 

The described phenomenological picture envisions turbulent eddies as magnetic current sheets with imposed sheared velocity profiles. The smaller the scale $a$, the more anisotropic the turbulent eddies become. It is however known that strongly anisotropic current sheets are subject to a fast-growing tearing instability and magnetic reconnection \cite[e.g.,][]{loureiro_instability_2007,uzdensky_magnetic_2014,pucci_reconnection_2014,comisso_general_2016,pucci2018}.  One can therefore ask whether the eddies formed by magnetic turbulence may become affected by the tearing instability, that is, whether the rate of the instability can become comparable to the eddy turnover rate at some small scale. This question was recently addressed in a series of works \cite[][]{loureiro2017,mallet2017,boldyrev_2017,comisso2018,walker2018,dong2018}, and it will be discussed in the next section.

\section{Tearing instability in MHD turbulence}
\label{tearing_mhd}
Consider an anisotropic turbulent eddy where a sheared magnetic field, that is, the fluctuating field in the $x-y$ plane, is approximated by $b_a f(x){\bf y}$. In our study of the tearing instability, we use the notation customary in the reconnection literature, where the thickness of the reconnection layer is denoted by $a$, while in the literature on turbulence, the thickness of a turbulent structure is usually denoted by~$\lambda$. We assume that the magnetic diffusivity is small but not negligible, and compare the resulting tearing rate with the eddy turnover rate. For that, we add a small perturbation to the magnetic field ${\bf b}=b_a f(x){\bf y}+\delta{\bf b}(x, y)$. We neglect the shearing profile of the velocity field, and take into account only the velocity perturbation, ${\bf v}=\delta{\bf v}(x, y)$. As argued in \citet{boldyrev_loureiro2018}, the velocity profile is not expected  to change the scaling of the fastest tearing growth rate, so we will not consider possible modifications caused by the shearing velocity profile our discussion. 

The function $f(x)$ describes the magnetic profile of an eddy. We will consider two model  profiles that lead to exactly solvable problems: $f(x)=\tanh(x/a)$ and $f(x)=\sin(x/a)$. In the first case, the magnetic field reverses its direction at a scale $a$ and it does not vary significantly outside the reversal region. In the second case, both the direction and the magnitude of the magnetic field vary at similar scales, which is probably a more common situation in a turbulent flow. In general, one can expect that the structures formed by turbulence lie between these two limiting cases. These two model profiles lead to slightly different scalings of the tearing modes, which, however, result in qualitatively similar solutions.  The perturbations of the magnetic and velocity fields can be expanded in Fourier series in the $y$ direction, $\delta b,\, \delta v \propto \exp(ik_0y)$. Then, the fastest growing Fourier mode (the so-called Coppi mode, e.g., \citet[][]{coppi1966,loureiro_instability_2007}) can be found from the solution of the linearized MHD equations. The analytical solution is rather involved, here we simply summarize the results for the fastest growing tearing mode (a detailed derivation can be found in, e .g., \citet[][]{boldyrev_loureiro2018}) for the $\tanh$- and $\sin$-profiles. In the case of the $\tanh$-shaped magnetic profile, the fastest growing tearing mode  has the wave number and the growth rate:
\begin{eqnarray}
k_{0, {\rm tanh}}\sim \eta^{1/4}b_a^{-1/4}a^{-5/4}, \label{k_tanh}\\
\gamma_{\rm tanh}\sim \eta^{1/2}b_a^{1/2}a^{-3/2}.\label{gamma_tanh}
\end{eqnarray}
In the sine-shaped profile, the tearing mode parameters are
\begin{eqnarray}
k_{0, {\rm sin}}\sim \eta^{1/7}b_a^{-1/7}a^{-8/7}, \label{k_sin}\\
\gamma_{\rm sin}\sim \eta^{3/7}b_a^{4/7}a^{-10/7}.\label{gamma_sin}
\end{eqnarray}
Here, $b_a$ is normalized to $\sqrt{4\pi \rho_0}$, so it is essentially the Alfv\'en velocity corresponding to the reconnecting field~$b_a$.

We now compare these rates with the eddy turnover rates of MHD turbulence. The magnitude of turbulent magnetic fluctuations at scale $a$ is estimated from the constant energy flux, $b_a^2 (b_a/a)(a/L)^{1/4}\sim \varepsilon $. Then, the nonlinear eddy turnover rate is estimated as
\begin{eqnarray}
\gamma_{nl}\sim (a/L)^{1/4} b_a/a \sim \varepsilon^{1/3} L^{-1/6} a^{-1/2}.\label{gamma_nl}
\end{eqnarray}
One can check that as scale $a$ decreases, the nonlinear eddy turnover rate increases slower  than any of the tearing rates (\ref{gamma_tanh}) or (\ref{gamma_sin}). Therefore, at some small scale, the tearing effects will become important, and the character of turbulence will change. For a tanh-shaped profile, the transition scale is 
\begin{eqnarray}
{a_c}/{L}\sim S^{-4/7},\label{ac_tanh}
\end{eqnarray}
while for a sine-shaped profile we get
\begin{eqnarray}
{a_c}/{L}\sim S^{-6/11},\label{ac_sine}
\end{eqnarray}
where $S=V_{A,L}L/\eta$ is the magnetic Reynolds number of a turbulent flow (or Lundquist  number as we assume that magnetic and velocity fluctuations are comparable at the outer scale $L$).  This number is assumed to be large in our considerations. 

\begin{figure}
\hskip10.mm\includegraphics[width=0.8\columnwidth]{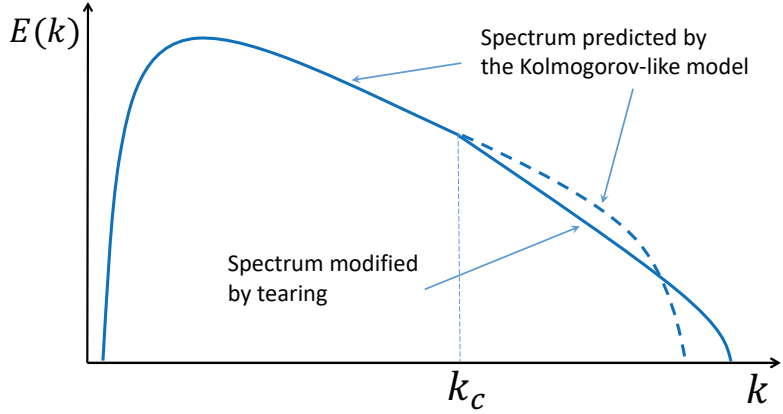}	
    \caption{Sketch of the Fourier energy spectrum of MHD turbulence mediated by the tearing instability. The tearing instability modifies the spectrum at scales smaller than the critical scale corresponding to~$k_c\sim 2\pi/a_c$. If the tearing instability were not taken into account, the spectrum at small scales would follow the Kolmogorov-like form, denoted by the dashed line.}
    \label{sketch_spectra}
\end{figure}

It is important to compare the scales associated with the tearing mode with the Kolmogorov-like dissipation scale of turbulence, which one would obtain if one did not take into account the tearing effects. The Kolmogorov-like resistive dissipation scale would be $a_\eta/L\sim S^{-2/3}$ (indeed, this is the scale where the eddy turnover rate (\ref{gamma_nl}) becomes comparable to the dissipation rate $\gamma_\eta\sim \eta/a^2$). We observe that in both cases, the tearing scale is larger than the would-be dissipation scale, meaning that the Kolmorogov-like dissipation mechanism will not be established in magnetic turbulence. The larger the Lundquist number, the larger the interval of scales affected by the tearing instability. The sketch in Figure~\ref{sketch_spectra} illustrates the modification of the energy spectrum of turbulence predicted by our model.

The tearing mode has another characteristic scale -- the so-called inner scale. In both model cases considered above, the inner scale corresponding to the fastest growing mode is estimated as $\delta_{\rm in} \sim a(\gamma/k_0b_a)$. One can verify that in both cases (\ref{gamma_sin}) and (\ref{gamma_tanh}), the inner scale is smaller than the would-be Kolmogorov dissipation scale  $a_\eta$, meaning that Kolmogorov-like turbulence would not destroy the inner structure of the tearing mode so that the mode can indeed develop in the current sheets formed by MHD turbulence.   

A natural question arising from our discussion is what is the spectrum of turbulence governed by the tearing instability at scales smaller than $a_c$. A possible answer was proposed in \cite[][]{loureiro2017,mallet2017,boldyrev_2017}, where it was suggested that the spectrum can be calculated by replacing the nonlinear turnover time by the tearing time at all scales smaller than $a_c$. As a result, for the tanh-like reconnection profile one gets $b_a^2\gamma_{\rm tanh}=\varepsilon$, which leads to the scaling of the fluctuating fields
\begin{eqnarray}
b_a\sim \varepsilon^{2/5}\eta^{-1/5}a^{3/5},
\end{eqnarray}
and to the energy spectrum
\begin{eqnarray}
E_{\rm tanh}(k_\perp) \propto k_\perp^{-11/5}.
\end{eqnarray}  
A similar estimate for the sine-shaped profile, $b_a^2\gamma_{\rm sin}=\varepsilon$, leads to  somewhat different scaling exponents
\begin{eqnarray}
b_a\sim \varepsilon^{7/18}\eta^{-1/6}a^{5/9},
\end{eqnarray}
and
\begin{eqnarray}
E_{\rm sin}(k_\perp)\propto k_\perp^{-19/9}.
\end{eqnarray}
Definitive numerical observations of tearing-mediated interval of turbulence require a rather large magnetic Reynolds number, exceeding $10^6$ as estimated by \citet[][]{boldyrev_2017}. Three-dimensional simulations would, therefore, be quite a challenging task and they are hardly possible with currently available computing power. Recent two-dimensional simulations of MHD turbulence performed by \citet[][]{walker2018} and \citet[][]{dong2018} seem to be able to detect the presence of tearing-mediated interval at small scales, yielding good agreement with the above predictions for the transition scale and the resulting scaling of the energy spectrum at scales smaller than that.

We would like to conclude this section with a historical note. To the best of our knowledge, the first analytic studies of the tearing effects in current sheets formed by MHD turbulence, date back to 1990, \citep{carbone1990}. That work employed the Iroshnikov \& Kraichnan (IK) model of MHD turbulence \citep{iroshnikov_turbulence_1963,kraichnan_inertial_1965}. The crucial aspect of the IK model is that MHD turbulence was assumed to be inherently weak, that is, consisting of essentially isotropic and weakly interacting Alfv\'en waves propagating along and against a background magnetic field. Based on this assumption, the IK analysis suggested that the spectrum of MHD turbulence should be~$-3/2$. A fundamental problem with the IK derivation was recognized about 30 years later, when it was realized that weak MHD turbulence, in fact, has to be strongly anisotropic with respect to the background magnetic field. A revisited theory of weak MHD turbulence demonstrated that the spectrum had to be $-2$ rather than $-3/2$ \cite[][]{goldreich1997,ng1996,galtier2000}. 

Our discussion, on the contrary, is devoted to strong MHD turbulence, where the $-3/2$ spectrum arises from the assumption of critical balance and dynamic alignment of magnetic and velocity fluctuations. The picture of strong MHD turbulence adopted in our work is fundamentally different from that of the IK model. However, as the \cite{carbone1990} consideration also assumed the $-3/2$ exponent of the turbulence spectrum, some of its scaling predictions, although based on an incorrect model of MHD turbulence, formally coincide with ours. For example, the critical scale predicted in \citet{carbone1990} coincides with our  Eq.~(\ref{ac_tanh}). For a more detailed discussion of this question, we refer the reader to \cite{boldyrev_loureiro2018}.

\section{Kinetic-Alfv\'en turbulence in a low electron-beta plasma}
\label{kaw_turbulence}

We now discuss a different regime of magnetic plasma turbulence, where particle collisions are so rare that the  magnetohydrodynamic resistive mechanism of tearing instability and energy dissipation is not relevant.  In such a collisionless plasma, the Alfv\'en turbulent cascade is not terminated by dissipation until it reaches scales smaller than the ion gyroscale. At such kinetic scales Alfv\'en turbulence is expected to transform into kinetic-Alfv\'en turbulence \cite[][]{howes_astrophysical_2006,howes08a,howes11a,boldyrev_p12,Groselj2018,roytershteyn2019,alexandrova09,kiyani09a,chen10b,chen12a,sahraoui13a,chen14b,boldyrev_etal2015}. In this section, we discuss the possible influence of the tearing instability on kinetic-Alfv\'en turbulence. We consider a collisionless plasma, and, similarly to the MHD case, assume that the fluctuations are strongly anisotropic with respect to the background magnetic field ${\bf B}_0=B_0{\hat z}$, so that $k_\perp\gg k_z$. We will also concentrate on the kinetic range of scales, $k_\perp\rho_i\gg 1$ (where $\rho_i$ is the ion gyroscale), and assume that the ion plasma beta, that is, the ratio of the ion kinetic energy to the energy of magnetic field, is of order one.  The magnetic {\em fluctuations} can then  be represented as ${\bf B}=-{\hat z}\times \nabla \psi+B_z{\hat z}$. In what follows we will use the dimensionless variables, ${\tilde \psi}=\psi/(d_e B_0)$ and $b_z=B_z/B_0$, where $d_e=c/\omega_{pe}$ is the electron inertial scale, and we will omit the tilde sign. Under these assumptions, the equations describing the sub-proton-scale kinetic-Alfv\'en fluctuations are \cite[][]{chen_boldyrev2017}:
\begin{eqnarray}
&\frac{1}{|\Omega_e|}\frac{\partial}{\partial t}\left(1-d_e^2\nabla_\perp^2\right)\psi - d_e^4\left[\left({\hat z}\times\nabla b_z\right)\cdot \nabla\right]\nabla_\perp^2\psi=-d_e\nabla_\|b_z,\label{equation1}\\
&\frac{1}{|\Omega_e|}\frac{\partial}{\partial t}\left(1+\frac{2}{\beta_i}-d_e^2\nabla_\perp^2\right)b_z - d_e^4\left[\left({\hat z}\times\nabla b_z\right)\cdot \nabla\right]\nabla_\perp^2 b_z=d_e^3\nabla_\|\nabla_\perp^2\psi.\label{equation2}
\end{eqnarray}
In these equations, $\Omega_e=eB_0/(m_e c)$ is the electron cyclotron frequency and the parallel gradient is the gradient along the instantaneous direction of the magnetic field, $\nabla_\|=\partial/\partial z - d_e \left({\hat z}\times \nabla \psi\right) \cdot \nabla $. The nonlinearities in this system enter through the second terms in the left-hand sides of the equations, and through the nonlinear term in the parallel gradient~$\nabla_\|$.

In the kinetic-Alfv\'en system (\ref{equation1}) and (\ref{equation2}) we kept the electron-inertia effects, which are described by the terms containing $d_e$ in the left-hand sides (note that the $d_e$ containing term in the parallel gradient, $\nabla_\|$, is {\em not} the inertial effect), but neglected the electron gyroscale terms.   This is a valid approximation when the electron beta is much smaller than one, $\beta_e=2\rho_e^2/d_e^2\ll 1$.  In this case, the tearing instability and magnetic reconnection that are governed by the electron-inertia effects, are separated from the electron-gyroscale, where  the electron Landau damping effects also come into play. In our discussion, we will therefore assume the smallness of the electron beta.  This will simplify the analytic treatment, and allow one to concentrate on the role of the  tearing effects. 

A more appropriate name for the considered equations is the ``inertial kinetic-Alfv\'en system'' since the electron inertia effects play an important role in the plasma dynamics \cite[e.g.,][]{roytershteyn2019,boldyrev2019}. Such a system is applicable to environments with low electron plasma beta, which include the solar corona, the Earth's magnetosheath, possibly collisionless shocks, etc. \cite[e.g.,][]{cranmer_etal2009,chandran_etal2011,ghavamian13,chen_boldyrev2017}. When necessary, however, the finite electron gyroradius terms can be taken into account\cite[see, e.g.,][]{passot2017,passot2018}. The system (\ref{equation1}),~(\ref{equation2}) is valid in the range of scales $\rho_e\ll a \ll \rho_i$. If, however, one is interested only in the scales exceeding the electron inertial  scale $d_e$, the system can be simplified by neglecting the terms containing $d_e$ in the left-hand sides of the equations. 

Similarly to our discussion of MHD dynamics, we now consider the conservation laws of the kinetic-Alfv\'en system. As one can directly verify, there are two second-order conserved integrals, which are the generalizations of the energy and magnetic helicity,
\begin{eqnarray}
 & E=\frac{1}{2}\int\left[b_z\left(1+\frac{2}{\beta_i}-d_e^2\nabla_\perp^2\right)b_z -d_e^2\nabla_\perp^2\psi\left(1-d_e^2\nabla_\perp^2\right)\psi \right] d^3 x,\label{e_kaw} \\   
& H^M=\int \left(1+\frac{2}{\beta_i}-d_e^2\nabla_\perp^2 \right)b_z \cdot \left(1-d_e^2\nabla_\perp^2\right)\psi \, d^3 x.\label{h_kaw}
\end{eqnarray}
As can be demonstrated, in kinetic-Alfv\'en turbulence $b_z\sim |\nabla_\perp \psi|$. At scales above the electron inertial scale, the small $d_e^2\nabla_\perp^2$ terms can be neglected in the parentheses of Eqs.~(\ref{e_kaw}) and~(\ref{h_kaw}). In this case, one can argue by  following the standard line of reasoning \cite[e.g.,][]{hasegawa1985,zakharov1992}, that since the energy involves higher-order gradients of the fields, it should cascade to small scales faster than the helicity.\footnote{Quite intriguingly, this trend has to reverse at scales smaller than $d_e$. Indeed, at $d_e^2 k_\perp^2\gg 1$, the higher-order gradients will be contained in the helicity rather than the energy integral. In this case, the helicity should cascade to small scales faster than the energy. The consequences of this type of selective decay are currently not fully understood. A possible scenario of a turbulent cascade in a similar case was proposed in \citet[][]{loureiro2018}.}

A standard derivation of the turbulent energy spectrum in this case goes as follows \cite[e.g.,][]{howes2008,boldyrev_p12}. According to Eq.~(\ref{e_kaw}), the energy density is estimated as $E_a\propto b_a^2$. The eddy turnover rate at the same scale can be found from, e.g., Eq.~(\ref{equation1}) as $\gamma_a\propto b_a/a^2$. The condition of constant energy flux $\gamma_a E_a\sim b_a^3/a^2\sim \varepsilon ={\rm const}$ then gives
\begin{eqnarray}
b_a\sim \varepsilon^{1/3}a^{2/3},
\end{eqnarray}
from which one can derive the Fourier energy spectrum as
\begin{eqnarray}
E(k_\perp)\propto k_\perp^{-7/3}.
\end{eqnarray}
Solar wind measurements of the turbulence spectrum at sub-proton scale, however, often produce a steeper spectrum, somewhat closer to $-8/3$ or $-3$~\cite[e.g.,][]{alexandrova09,kiyani09a,chen10b,chen12a,sahraoui13a,chen2016}. Various explanations have been proposed for the steepening of the spectrum with respect to this dimensional estimate; some of them invoke Landau damping effects and some intermittency corrections \cite[][]{howes2008,boldyrev_p12,tenbarge2012}. In what follows we argue that the tearing instability of highly anisotropic turbulent eddies and resulting tearing-mediated kinetic-Alfv\'en turbulence may provide an alternative and complementary explanation for the observed spectral steepening.  

First, we note that at scales smaller that $d_e$, magnetic field is not frozen into the fluid and reconnection cannot take place. We will, therefore, be interested in magnetic configurations at scales larger than~$d_e$. In order to minimize the integral of energy keeping the magnetic helicity constant, we need to minimize the functional $W=E-\mu H^M$, where $\mu$ is a Lagrange multiplier. Varying the functional with respect to $\psi$ and $b_z$, and assuming that $k_\perp d_e\ll 1$ we obtain, correspondingly,
\begin{eqnarray}
d_e^2\nabla_\perp^2\psi=-\mu b_z, \label{constraint1}\\
b_z=\mu \left(1+2/\beta_i \right)^{-1}\psi.\label{constraint2}
\end{eqnarray}
These equations provide us with two important pieces of information. First, they tell us what configurations of the fields minimize the energy under the constraint of constant helicity.  Second, they show that the nonlinear interactions in the kinetic-Alfv\'en system~(\ref{equation1}) and (\ref{equation2}) corresponding to these configurations, are identically zero. We see that the parameter $\mu$ plays the role of the inverse scale of the corresponding configuration. If we substitute expressions (\ref{constraint1}) and  (\ref{constraint2}) in the helicity and energy integrals, we easily obtain that $H^M=\int b_z\psi d^3 x=\mu \left(1+2/\beta_i \right)^{-1} \int \psi^2 d^3 x$, and $E=\mu^2 \left(1+2/\beta_i \right)^{-1}\int \psi^2 d^3 x$, from which it follows that $E=\mu H^M$.  Therefore, in order to minimize the energy keeping the helicity constant, we need to decrease the scale of the configuration. If we were dealing with the case of non-driven (decaying) turbulence, this would imply that helical magnetic structures would tend to form at large scales.   

However, our goal is to describe steady-state turbulence, which is governed by a constant energy flux from large scales where turbulence is generated, toward small scales where turbulent energy is converted into internal energy of plasma particles. The extreme configurations (\ref{constraint1}) and (\ref{constraint2}) cannot be formed in this case, since  the nonlinear interaction responsible for the energy flux cannot vanish. It may, however, be reasonable to expect that the formed structures tend to satisfy the conditions (\ref{constraint1}, \ref{constraint2}) approximately and that they also tend to maximize the helicity {\em given a constant energy flux}. The latter requirement is rather nontrivial, as the standard statistical methods do not apply to open turbulent systems. Formulated in a somewhat simplified way, our requirement may suggest that the turbulent dynamics tends to minimize the nonlinear interaction for given amplitudes of the fields. (This is qualitatively similar to our discussion of MHD turbulence, which may possibly suggest that spontaneous formation of structures where the nonlinearity is reduced is a general property of magnetic plasma turbulence.) 

How can this be achieved? Let us assume that the structures have typical wavenumbers $k_x$ and $k_y$ in the field-perpendicular directions. From Eqs.~(\ref{constraint1}, \ref{constraint2}) we then estimate that $b_z\sim \mu \psi$ and $\nabla_\perp^2\psi\sim \mu^2\psi$, where the scale of the structure is characterized by $\mu^2\sim k_x^2+k_y^2$.   The nonlinear terms in Eqs.~(\ref{equation1}, \ref{equation2}) are then estimated for these structures as $k_xk_y\psi^2$, while the helicity is proportional to $\mu \psi^2$. If the helicity is constant and the scale of the structure, $\mu$, is given, then the nonlinear terms become small when only one of the wavevector components, $k_x$ or $k_y$, becomes small. This condition is consistent with the formation of anisotropic  current-sheet structures. The current sheets may, however, become unstable to the tearing mode, which can arrest their formation and modify the character of kinetic-Alfv\'en turbulence at the corresponding scales. We discuss this question in the next section.

\section{Tearing instability in kinetic-Alfv\'en turbulence}
\label{tearing_kaw}

Consider an anisotropic current sheet (a turbulent eddy) with a thickness $a$, where the  magnetic field is given by ${\bf b}=b_a f(x){\bf y}$. The magnetic profile described by the function $f(x)$ has a characteristic scale $a$. As in the case of MHD turbulence, we will consider two analytically solvable model profiles, $f(x)=\tanh(x/a)$ and $f(x)=\sin(x/a)$, keeping in mind that realistic magnetic profiles formed in turbulence probably fall in between these two limiting cases.  Similarly to the magnetohydrodynamic case, we investigate stability of the current sheet with respect to the tearing mode by adding a small perturbation, ${\bf b}=b_a f(x){\bf y} +\delta{\bf b}(x, y)$. A remarkable fact that has been stressed in e.~g., \citet[][]{boldyrev2019} is that the linearized equations describing the tearing mode in the kinetic-Alfv\'en case are structurally analogous to those  derived in the magnetohydrodynamic case. The only difference is that the kinetic-Alfv\'en equations (\ref{equation1}, \ref{equation2}) do not have resistive dissipation, and, threrefore, the flux freezing condition is broken not by the magnetic diffusivity but rather by the electron inertia, that is, the term $(\partial/\partial t)d_e^2\nabla_\perp^2\psi$ in the left-hand side of Eq.~(\ref{equation1}). The tearing mode in this case can be obtained from the equations derived for the MHD case, Eqs.~(\ref{k_tanh}, \ref{gamma_tanh}) or Eqs.~(\ref{k_sin}, \ref{gamma_sin}), if one replaces $\eta\to \gamma d_e^2$. 
As a result, we obtain the following characteristic parameters of the fastest growing tearing mode for the tanh-shaped magnetic profile:
\begin{eqnarray}
&k_{0, {\rm tanh}}\sim d_e/a^2, \label{k_kaw_tanh}\\
&\gamma_{\rm tanh}\sim b_ad_e^2/a^3,
\end{eqnarray}
while for the sine-shaped magnetic profile we analogously find
\begin{eqnarray}
&k_{0, {\rm sin}}\sim d_e^{1/2}/a^{3/2}, \label{k_kaw_sin}\\
&\gamma_{\rm sin}\sim b_ad_e^{3/2}/a^{5/2}.
\end{eqnarray}
In both cases, the corresponding inner scales of the mode are $\delta_{\rm in}\sim d_e$. 

We therefore envision the following picture of kinetic-Alfv\'en turbulence. Turbulent dynamics favor the formation of eddies that have a morphology of anisotropic current sheets in the inertial interval. However, strongly anisotropic eddies become susceptible to the tearing mode, which sets the limit on the anisotropy that can be achieved. As the  current sheet becomes progressively more anisotropic, that is, as $k_0$ decreases, the corresponding tearing rate increases while its nonlinear turn-over rate decreases (for a detailed analysis, we refer the reader to \cite{boldyrev2019}). One can demonstrate that, quite interestingly, the tearing rate overcomes the turnover rate at exactly the same scale where the tearing rate reaches its maximal value, that is, the value given by expressions (\ref{k_kaw_tanh}) or (\ref{k_kaw_sin}). The corresponding scale then defines the anisotropy of the turbulent structures. 

We now discuss the spectra of the resulting tearing-mediated kinetic-Alfv\'en turbulence.  According to the picture described above, anisotropic structures formed by turbulence are marginally tearing-unstable at all scales, therefore, their turn-over rate is on the order of the tearing rate. Analogously to the MHD case, we then estimate the energy flux as $b_a^2\gamma_{\rm tanh}\sim \varepsilon={\rm const}$ or $b_a^2\gamma_{\rm sin}\sim \varepsilon={\rm const}$ depending on the assumed magnetic profile, which gives for the tanh-shaped profile:
\begin{eqnarray} 
b_a\sim \varepsilon^{1/3}a^{1/3}\left(a/d_e\right)^{2/3},
\end{eqnarray}
while for the sin-shaped profile we obtain
\begin{eqnarray}
b_a\sim \varepsilon^{1/3}a^{1/3}\left(a/d_e\right)^{1/2}.
\end{eqnarray}
The corresponding Fourier energy spectra scale as
\begin{eqnarray}
E_{\rm tanh}(k_\perp)\propto k_\perp^{-3},\\
E_{\rm sin}(k_\perp)\propto k_\perp^{-8/3}.
\end{eqnarray}
Spectra that are roughly consistent with these predictions are indeed detected in {\it in situ} measurements of space plasma turbulence. Also, the effects associated with the ``electron-only'' reconnection events have been recently detected in turbulence in the Earth's magnetosheath (with the NASA MMS mission, \citet{phan2018}), which is consistent with our predictions.  The parameters ordering assumed in our model, $\beta_e\ll \beta_i\sim 1 $, is especially relevant for the vicinity of the solar corona that will soon be probed by the NASA Parker Solar Probe mission \cite[][]{bale2016}. Our theory of tearing-mediated kinetic-Alfv\'en turbulence may therefore provide a plausible analytical framework for these studies.

\section{Conclusion}
\label{conclusions}
The goal of our discussion was to give a qualitative introduction to the theory of tearing-mediated magnetic plasma turbulence. In recent years, it became clear that the tearing instability may significantly affect the small-scale behavior of magnetic turbulence. In particular, inertial-range-scale current sheets that have often been envisioned as structures formed in a turbulent cascade, are expected to become unstable to the tearing mode, and alter the cascade of energy to smaller scales. 

In the applications considered, tearing effects have been found to come into play before the would-be Kolmogorov dissipation scale is reached by an energy cascade. The larger the Reynolds number of turbulence, the larger the range of scales altered by the tearing effects. The importance of  this phenomenon is, however, not only in modifying the spectra of turbulence. Understanding the structure of small-scale fluctuations may also be important for understanding the route of energy dissipation in magnetic plasma turbulence, for instance, the relative importance of particle acceleration with respect to a thermal deposition of turbulent energy.

In the present work, we have illustrated how the energy spectrum is expected to be modified by the tearing instability in two model cases, the resistive magnetohydrodynamics and the collisionless  kinetic-Alfv\'en dynamics. Of course, this discussion does not exhaust all the possible cases of tearing-mediated plasma turbulence. Our results for the inertial kinetic-Alfv\'en case can be extended almost straightforwardly to whistler turbulence (described by the electron MHD). Similar theories have also been recently developed for collisionless Alfv\'en turbulence and for pair-plasma turbulence \cite[][]{loureiro2018,loureiro2017a}, and they can possibly be extended to other plasma phenomena where turbulence plays an important role.

\acknowledgments
The work of SB was partly supported by the NSF under grant no. NSF PHY-1707272, by NASA under grant no. NASA 80NSSC18K0646, and by DOE grant No. DE-SC0018266. NFL was partially funded by NSF CAREER award no. 1654168 and by the NSF-DOE Partnership in Basic Plasma Science and Engineering, award no. DE-SC0016215.


%


%




\end{document}